\documentclass[sigconf]{acmart}

\usepackage{amsmath}
\usepackage{bookmark}
\usepackage{xspace}

\usepackage{paralist}

\usepackage[htt]{hyphenat}

\usepackage{hyperref}
\hypersetup{colorlinks=true,urlcolor=black,hidelinks,breaklinks=true}

\usepackage{xurl}
\usepackage[table,xcdraw]{xcolor}

\usepackage{mathtools}

\usepackage{listings}

\usepackage{enumitem}

\newif \iftopics \topicsfalse
\iftopics
\newcommand{\topic}[1]{{\textcolor{gray}{[#1]\\}}}
\newcommand{\subtopic}[1]{{\textcolor{gray}{[#1]\\}}}
\else
\newcommand{\topic}[1]{}
\newcommand{\subtopic}[1]{}
\fi

\usepackage{prettyref}

\newcommand{\figreffmt}[1]{Figure~\ref{#1}}
\newcommand{\tblreffmt}[1]{Table~\ref{#1}}
\newcommand{\secreffmt}[1]{Section~\ref{#1}}

\newrefformat{fig}{\figreffmt{#1}}
\newrefformat{tab}{\tblreffmt{#1}}
\newrefformat{tbl}{\tblreffmt{#1}}
\newrefformat{sec}{\secreffmt{#1}}
\newrefformat{subsec}{\secreffmt{#1}}
\newrefformat{subsubsec}{\secreffmt{#1}}

\newcommand{\ie}{\textit{i.e.}\xspace}
\newcommand{\eg}{\textit{e.g.}\xspace}

\newcommand{\etc}{\textit{etc.}\xspace}

\usepackage{relsize}

\usepackage{scalerel}

\newcommand{\doubleplus}{{+\kern-1.3ex+\kern0.8ex}}

\usepackage{amsfonts}

\definecolor{mymauve}{rgb}{0.58,0,0.82}
\definecolor{CommentGreen}{rgb}{0.58,0,0.82}

\definecolor{SolarizedBase03}{HTML}{002b36}
\definecolor{SolarizedBase02}{HTML}{073642}
\definecolor{SolarizedBase01}{HTML}{586e75}
\definecolor{SolarizedBase00}{HTML}{657b83}
\definecolor{SolarizedBase0}{HTML}{839496}
\definecolor{SolarizedBase1}{HTML}{93a1a1}
\definecolor{SolarizedBase2}{HTML}{eee8d5}
\definecolor{SolarizedBase3}{HTML}{fdf6e3}
\definecolor{SolarizedYellow}{HTML}{b58900}
\definecolor{SolarizedOrange}{HTML}{cb4b16}
\definecolor{SolarizedRed}{HTML}{dc322f}
\definecolor{SolarizedMagenta}{HTML}{d33682}
\definecolor{SolarizedViolet}{HTML}{6c71c4}
\definecolor{SolarizedBlue}{HTML}{268bd2}
\definecolor{SolarizedCyan}{HTML}{2aa198}
\definecolor{SolarizedGreen}{HTML}{859900}

\renewcommand\footnotetextcopyrightpermission[1]{}

\begin{document}

\newcommand\boxedtext[1]{
    \vspace{0.25\baselineskip}
	\noindent\fbox{
        \parbox{\linewidth - 3\fboxsep}{#1}%
	}\vspace{0.1\baselineskip}
}

\newcommand{\ringleader}{Ringmaster\xspace}

\title{\ringleader: How to juggle high-throughput host OS system calls from TrustZone TEEs}

\author{Richard Habeeb}
\affiliation{%
    \institution{Yale University}
    \city{New Haven}
    \state{Connecticut}
    \country{USA}
}

\author{Man-Ki Yoon}
\affiliation{%
    \institution{North Carolina State University}
    \city{Raleigh}
    \state{North Carolina}
    \country{USA}
}

\author{Hao Chen}
\affiliation{
    \institution{Yale University}
    \city{New Haven}
    \state{Connecticut}
    \country{USA}
}

\author{Zhong Shao}
\affiliation{
    \institution{Yale University}
    \city{New Haven}
    \state{Connecticut}
    \country{USA}
}

\begin{abstract}
Many safety-critical systems require timely processing of sensor inputs to
avoid potential safety hazards. 
Additionally, to support useful application features, such systems increasingly
have a large rich operating system (OS) at the cost of potential security bugs.
Thus, if a malicious party gains supervisor privileges, they could cause
real-world damage by denying service to time-sensitive programs.
Many past approaches to this problem completely isolate time-sensitive programs
with a hypervisor; however, this prevents the programs from accessing useful OS
services. 
We introduce \ringleader, a novel framework that enables enclaves or TEEs
(Trusted Execution Environments) to asynchronously access rich, but potentially
untrusted, OS services via Linux’s \textit{io\_uring}.
When service is denied by the untrusted OS, enclaves continue to operate on
\ringleader's minimal ARM TrustZone kernel with access to small, critical device
drivers.
This approach balances the need for secure, time-sensitive processing with the
convenience of rich OS services.
Additionally, \ringleader supports large unmodified programs as enclaves,
offering lower overhead compared to existing systems.
We demonstrate how \ringleader helps us build a working highly-secure system
with minimal engineering.
In our experiments with an unmanned aerial vehicle, \ringleader achieved nearly
1GiB/sec of data into enclave on a Raspberry Pi4b, 0-3\% throughput
overhead compared to non-enclave tasks.

\end{abstract}
\maketitle

\pagestyle{plain} %

\section{Introduction}
\label{section:intro}
Recent advancements in computer vision and artificial intelligence (AI) have
driven significant progress in autonomous vehicles, drone applications, surgical
robotics, and other safety-critical cyber-phys\-ical systems (CPS).
However, ensuring the safety, reliability, and security of these systems remains
a critical challenge \cite{SecurityModernAutomobile2010,
ComprehensiveAutomotiveAttack2011, SurgicalRobotsAnalysis2015,
PreparingForAutonomousVehicles2015, IndustrialRobotControllerAnalysis2017,
SoKIoTCyberattack2018, DroneSecurityAnalysis2020, AVCybersecurityReview2021,
SoKDroneSecurity2021, SoKIndustrialControlSystemsSecurity2020}.
This stems from the need for rich operating systems and complex software stacks
to support machine learning (ML) models and features like live video streaming
and voice recognition.
The Linux kernel, despite its maturity, still contains security vulnerabilities
\cite{LinuxCVETracker}, made worse by third-party libraries, drivers, and
packages, which expand the attack surface.
Additionally, modern CPS are often internet-connected for remote operations and
live data streaming \cite{TheAirGap2013, AirGapIsDead}, increasing exposure to
threats.
These systems frequently use system-on-a-chip (SoC) designs \cite{NVIDIAThor,
MixedCriticality2022, ShiftToMulticores2015}, co-locating
\textit{time-sensitive} programs with less critical ones, which introduces new
security risks \cite{SoKRealTimeSecurity2024}.
Despite prior research efforts, attacks continue to emerge
\cite{StuxNet2011, UkrainianPowerAttack2016, HackingAProfessionalDrone2016,
FreeFallHackingTesla2017, TheConnectedCar2018, RoadwaysToExploitBMW2019,
TBONE2020, DJIDroneSecurity2023, DownedUSDrone2023, RCETeslaInfotainment2024,
JailbreakingAnElectricVehicle2023, IFeelADraft2022}.

Trusted Execution Environments (TEEs) and \textit{enclaves} (in this paper we
use these terms interchangeably) have been explored to isolate software from a
privileged host operating system \cite{TrustedExecutionEnvironment2015}, but
they are not widely used for time-sensitive applications in practice.
An ideal CPS enclave might serve as a flight controller or collision-avoidance
system, for instance, ensuring \textit{availability} to respond to sensor inputs
promptly while communicating with remote operators or logging data to a disk.
Running the controller in an enclave would drastically reduce the
trusted-computing base (TCB), protecting critical software against
privilege-escalation attacks.
Focusing on the ARM TrustZone \cite{TrustZone2009}, despite many years of
quality research on TEEs \cite{DemystifyingARMTrustZone2019}, with numerous
strong defense designs for real-time systems \cite{SafeG2010,
DualOSScheduling2012, DualOSCommunication2013, SASP2013, RTZVisor2016,
LTZVisor2017, microRTZVisor2017, multicoreLTZVisor2017, VOSYSMonitor2017,
RT-trust2018, TZ-VirtIO2018, TZDKS2018, TZMVirt2019, PSpSys2022, RT-TEE2022,
uTango2022}; it is difficult to find real-world examples of them being used to
ensure availability in practice.
From our observations, TEEs for deployed CPS appear to largely be used for
secure boot or cryptographic operations \cite{DJIWhitePaper2024}.

At a high level, we believe there are at least two major reasons for the general
lack of enclave use in practice for time-sensitive applications.
First, the most practical designs---those which support untrusted OS system
calls, POSIX-compliance, or even unmodified applications---are not designed for
availability.
Second, the designs which focus on availability can be difficult to use; they
lack comprehensive OS services and have a highly limited programming
environment.
We expand below.

\paragraph{Enclaves with rich OS support face timing vulnerabilities}
Existing solutions that support rich OS services \cite{Proxos2006,
Overshadow2008, InkTag2013, VirtualGhost2014, MiniBox2014, Haven2014, Haven2015,
SCONE2016, Eleos2017, TrustShadow2017, Graphene-SGX2017, Panoply2017,
Occlum2020, CHANCEL2021, BlackBox2022} completely rely on the untrusted OS for
scheduling and memory management, making them unsuitable for time-sensitive
applications.
For instance, an adversary could delay page fault handling to subtly manipulate
the response time of a program; or it could simply kill the process.
Furthermore, the traditional, ``blocking'' POSIX-like programming model for
system calls gives the adversary full control over enclave timing.
An adversary could delay the return of an \texttt{open()}, \texttt{write()}, or
\texttt{read()} system call to affect the response time of some critical
behavior.

\paragraph{Enclaves with availability protection lack comprehensive OS support}
Designs with availability protections \cite{SafeG2010, DualOSScheduling2012,
DualOSCommunication2013, SASP2013, RTZVisor2016, LTZVisor2017,
microRTZVisor2017, multicoreLTZVisor2017, VOSYSMonitor2017, HYDRA2017, SMACCM,
SMACCM2017, RT-trust2018, TZ-VirtIO2018, TZDKS2018, TZMVirt2019, Aion2021,
ERTOS2021, PSpSys2022, RT-TEE2022, uTango2022, GAROTA2022} all have very limited
OS services.
This limits the scope of enclave functionality.
By design, ARM TrustZone enclaves \cite{TrustZone2009,
DemystifyingARMTrustZone2019} only have minimal Secure-world OS services because
the trusted computing base (TCB) would have to be inflated to implement a file
system or network stack, for example.
Additionally, limited system call support forces developers to implement custom
data transfer protocols \cite{VirtIO2008, TZ-VirtIO2018, RealTimeVirtio2021,
PikeOSVirtIO, SMACCM2017, Bao2020}.
Thus, small changes to tasks require substantial effort, making this approach
unappealing in practice.

\textbf{We propose \textit{\ringleader,}} a new design for making host OS system
calls from the TrustZone without sacrificing availability, real-time, or
performance needs.
\ringleader leverages the \texttt{io\_uring} framework
\cite{RingingInIOURing2019, RapidGrowthIOURing2020} for asynchronous system
calls, decoupling enclave timing from the system call protocol itself.
This approach provides much-needed, practical access to host OS services from
TrustZone enclaves (when the OS is well-behaved), and it prevents an enclave
from \textit{unnecessarily} blocking if the OS maliciously delays the result.
\ringleader's design unlocks the ability to easily communicate over encrypted
network channels from an enclave, to write encrypted data to a disk, or to
perform essential inter-process communication (IPC) through pipes and standard
I/O---all of which previously required a great degree of manual effort for
TrustZone TEEs.
This approach further unlocks the ability to run unmodified software in TEEs,
allowing the TrustZone to be used as a transparent security layer, whereas
existing TrustZone solutions all require significant security expertise and
manually engineered custom ``trustlets.''

Since an adversarial OS may choose to never ``return'' (even for an asynchronous
call), time-sensitive I/O is routed through a \ringleader-owned device via a
separate interface.
In practice, we observe such device drivers are often small (\eg{} UART, I2C,
SPI, or CAN) and we found robustly tested open-source versions available.
\ringleader ensures that enclaves will not starve by preempting and scheduling
Linux with a real-time scheduler.
We implemented \ringleader on the PARTEE real-time ARM TrustZone OS
\cite{PARTEE2025}, however, the design likely generalizes to other hardware TEE
platforms or isolation primitives.

\begin{figure}[t]
	\centering
	\includegraphics[width=1.0\columnwidth]{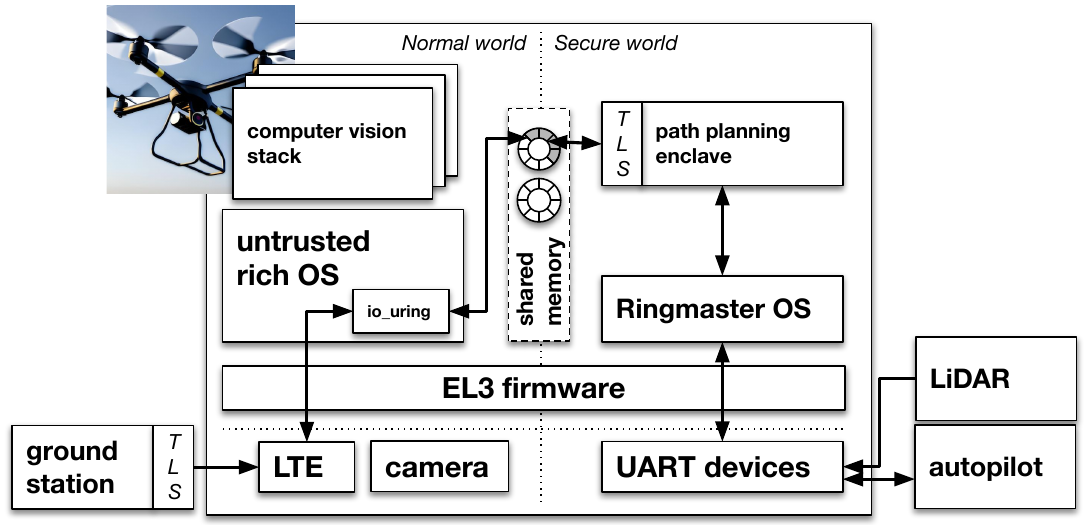}
	\caption{Example of \ringleader on a drone with a flight controller enclave
	communicating over encrypted asynchronous \texttt{io\_uring} operations with a
	remote operator; if the untrusted OS denies network service, the enclave
	will continue to stabilize and direct the drone.}
	\label{fig:ringleader-drone-overview}
\end{figure}

Without \ringleader, even basic needs, like network I/O, require overly-complex
engineering to get data into the TEE apps.
Consider the autonomous quadcopter (Fig. \ref{fig:ringleader-drone-overview}),
that we built to evaluate \ringleader.
Using our new approach, \ringleader can run a ground-station facing network
server directly in the TEE, which simply uses Linux network sockets.
Such a program can be written with minimal development effort.
Expanding on this, we run other critical software in the enclave as well, so
that even if high-level objectives fail to arrive, the enclave can
still direct the system's autopilot over serial to ensure safe behavior.
With encryption, the system detects tampering of network traffic, ensuring
robust operation even under adversarial conditions.
The enclave can send detailed flight logs and raw sensor data back to the
ground-control station with high-throughput, due to the parallelism of
\texttt{io\_uring} and \ringleader's zero-copy argument passing scheme.

The number of use cases for \ringleader is potentially quite large, especially
for  time-sensitive internet-of-things (IoT) or CPS.
For example, a smart traffic-management system could optimize signal timing with
live traffic data coming from the network, but it should still remain
operational and timely if incoming data is denied.
An AI-enhanced medical device should have clear isolation from timing-critical
medical sensing and actuation while also being able to both connect to the cloud
and to record sensor data into the file system.
Furthermore, our above drone could include a computer-vision Linux program which
detects fires or stranded hikers.
Even if the application is adversarial, it can only send high-level objectives
to the flight controller, greatly limiting the impact of the attack.
Because many of these systems have no security against this kind of attack,
\ringleader significantly reduces the TCB for critical tasks.

Legacy POSIX (Portable Operating System Interface) applications can benefit from
\ringleader as well.
Using timeouts and signals, such a program can be minimally updated to protect
it from being blocked waiting infinitely.
We built a custom \ringleader LibC that allows many unmodified programs to run
in the TrustZone---which has only been achieved by one other work
\cite{TrustShadow2017}.

\vspace{0.75\baselineskip}
\noindent\textbf{\textit{Contributions:}}
\begin{itemize}
    \item We design a new approach for initiating Linux system calls from
    independently managed enclaves with a zero-copy mechanism for transferring
    large arguments (\S \ref{section:syscall-design}).
    This includes handling of power and thermal management, mitigating the
    potential impacts of polling (\S \ref{section:power}).

    \item We present an asynchronous programming model that enables
    time-sensitive enclaves to securely request potentially untrusted services
    from the host OS, including an optimized shared memory manager (\S
    \ref{section:arena}).

    \item We demonstrate a design that supports minimally modified POSIX applications with non-starvation guarantees (\S \ref{section:ringleader-libc}).
    Experimental results demonstrate comparable or better latency for system
    calls than past approaches for unmodified enclaves (\S
    \ref{section:unmodified-performance}).

    \item We deliver a prototype implementation on the Raspberry Pi4B, a demonstration on a drone platform (\S \ref{section:drone-implementation}), and evaluations using unmodified GNU Core Utilities (\S \ref{section:unmodified-performance}).
    We showcase high throughput TEE I/O, with our experimental configuration fully saturating the Pi’s Ethernet port and nearly 1GiB/s sequential buffered  file system read and write speeds.
\end{itemize}

\section{Untrusted System Calls for Time-Sensitive Enclaves?}
\label{section:drone-motivation}
The intersection of three properties poses significant challenges for many CPS:
(1) \textbf{time sensitivity}, requiring predictable response times to events
(ranging from hard real-time deadlines to softer requirements without worst-case
execution time analysis);
(2) \textbf{distrust of privileged software}, which might be compromised; and
(3) \textbf{reliance on rich OS services}, such as networking, file systems,
pipes, and drivers, as well as third-party software.

While it may seem paradoxical to seek services from an untrusted OS, this is
common in modern CPS, IoT devices, and robots.
For example, drones often use the MAVLink protocol \cite{MAVLinkSurvey2019} to
transmit waypoints from ground stations or companion computers
\cite{TX2Companion, PixHawkCompanion} to real-time autopilot systems like
ArduPilot \cite{ArduPilot}, which stabilize the drone.
Because radio links are inherently unreliable, drones prioritize immediate
sensor data over MAVLink packet timing.

Similarly, autonomous vehicles use remote teleoperation in unexpected situations
(\eg{}, construction sites) for high-level guidance \cite{ZooxTeleGuidance2020} while
continuing real-time obstacle avoidance.
For instance, a number of recent security analyses
\cite{FreeFallHackingTesla2017, TBONE2020, IFeelADraft2022,
JailbreakingAnElectricVehicle2023, RCETeslaInfotainment2024} reveal that the Tesla
Autopilot system is directly connected to an Ethernet switch with internet
connectivity.

In both cases, time-sensitive systems depend on OS services (\eg{} from
Automotive Grade Linux \cite{AutomotiveGradeLinux} or Real-Time Ubuntu
\cite{RealTimeUbuntuWhitePaper2023} for instance), such as networking, while
maintaining robust real-time performance.
Other examples include writing logs to the file system, or inter-process
communication (IPC) between less critical and more critical tasks.

To meet this need, CPS SoC platforms must enable secure access to rich OS
services without fully trusting the OS.
\ringleader addresses this challenge by significantly reducing the trust placed
on the OS while maintaining the functionality required for time-sensitive tasks.
Our approach enables safe operation even when the OS is compromised.
We discuss approaches to handling malicious data in
\S\ref{section:malicious-io}, and comparisons with other related work in
\S\ref{section:related-work}.

\section{Models \& Assumptions}
Our models for enclaves, platforms, and adversaries build upon Subramanyan et
al.'s definitions \cite{FormalFoundationSRE2017}, extending them to address
time-sensitivity and availability.

\subsection{Hardware Model}
\label{section:hardware-model}
This work targets hardware platforms that provide privilege levels above kernel
mode with a memory-management unit, although it is not tied to a specific
architecture.
The hardware requirements include:
\begin{itemize}

	\item \textbf{Memory sharing and isolation.} Programmable control of OS memory
	access, for example, a TrustZone Address-Space Controller (TZASC) \cite{TZASC},
	RISC-V's physical-memory protection (PMP), or nested page tables (NPT).%
    \footnote{This potentially includes ARMv9's Granule Protection Table.
    \cite{ARMCCAManual}.}

	\item \textbf{Device isolation.} Configurable OS access to memory-map\-ped I/O
	(MMIO) and secure interrupt delivery \eg{} using ARM's Generic Interrupt
	Controller (GIC)\cite{GICManual}.

	\item \textbf{A dedicated timer.} An unmaskable timer interrupt for preemption
	and scheduling of the OS and enclaves.

	\item \textbf{Power, voltage, frequency isolation.} Independent management of
	system power, clock speed, and voltage.

\end{itemize}

Intel SGX, as it stands, does not match this model due to its reliance on kernel
mode for device and interrupt handling, and lack of programmable enclave page
table management.
Though our implementation is based on the ARMv8-A TrustZone, a hypervisor-based
isolation approach, \eg{} SMACCM \cite{SMACCM, SMACCM2017}, or a RISC-V
approach, as used by Keystone and ERTOS \cite{Keystone2020, ERTOS2021}, also
fits our model.

\subsection{Adversary Model}
\label{section:adversary-model}
\newcommand{\alice}{$\mathcal{A}$\xspace}
\newcommand{\bob}{$\mathcal{B}$\xspace}
\newcommand{\eve}{$\mathcal{E}$\xspace}

We define the adversary, \textit{Eve} (\eve), through two security games
addressing real-time guarantees and traditional enclave integrity and
confidentiality.
In both games, \eve has access to kernel-mode privileges and can read or write
to any physical address that is not protected by the mechanisms defined above.
\eve may schedule any process, and configure any interrupt, but she can also be
interrupted by the enclave platform.

\textbf{Game 1: Timeliness.}
Suppose a victim enclave, \textit{Alice} (\alice), must write a valid message to
a serial device, in a tight time bound $\mathcal{T}$.
However, the message is stored by \eve.
Upon request, \eve chooses to send either a valid or invalid message.
\alice has a function $f(m)$ that decides if a message is valid or not, if it is
invalid she can send a valid \textit{back-up} message.
Without any interference or starvation, \alice will check and write a valid
message in $\mathcal{T}/2$ time, even with the maximum number of cache misses,
page faults, branch mispredictions, or other microarchitectural delays.
In this game, the challenger is an enclave platform, and to win it must provide
a strategy to transfer the message to \alice such that: if \eve returns a valid
message before $\mathcal{T}/2$, \alice will write it.
Timely but incorrect responses are addressed under Game 2.

\textbf{Game 2: Integrity \& Confidentiality.}
Suppose our user process from Game 1, \alice, wants to receive a secret
from an external party, \textit{Bob} (\bob).
\eve can inspect messages, deny message delivery, modify or delete files,
or perform or deny any other kernel action.
\eve wins the game if she can read or modify communication between \alice and
\bob without their knowledge.
\eve also wins if she can read or infer secrets from \alice's memory.

\subsection{Assumptions}
\textbf{Responses to rich I/O starvation or corruption are out of scope.}
Lack of data is a problem, and solutions are ultimately application-specific.
The power of \ringleader is that \textit{enclaves can respond} to this
scenario.
Additionally, even without guaranteed availability for OS services, having
access is often still advantageous (see \S \ref{section:drone-motivation}).
Likewise, malicious I/O should be handled via encryption, authentication, or an
application-specific method.
We discuss further solutions in \S\ref{section:malicious-io}.

\textbf{Physical and microarchitectural attacks are out of scope.}
We assume that the adversary does not have physical access to the SoC, and
cannot sniff the memory or device buses, perform cold-boot attacks
\cite{ColdBoot2009}, power-analysis attacks
\cite{DifferentialPowerAnalysis1999}, \etc{}.
Microarchitectural side-channel attacks like ARMageddon, \etc{}
\cite{ARMageddon2016, TruSense2018}, are handled using emerging techniques
\cite{SecTEE2019, SassCache2023}.
Additionally, we assume hardware contentions---adversarial cache accesses,
page-table walks, branch predictions, memory bandwidth consumption, \etc{}---are
being addressed with complementary work \cite{PALLOC2014, MemGuard2013,
MemPol2023, RealTimeVirtio2021, Bao2020}.

\textbf{Implementation correctness.}
We assume that the business logic of \alice is written correctly.
We trust (and test) that our implementation of \ringleader is written
correctly.
Finally, we trust our build environment and compiler.

\section{Asynchronous System Call Design}
\label{section:syscall-design}
Here we discuss \ringleader's novel design for making arbitrary I/O system
calls into Linux from an enclave.

\paragraph{Background on \texttt{io\_uring}.}
Modern Linux kernels have an interesting system called \texttt{io\_uring}
\cite{RingingInIOURing2019, RapidGrowthIOURing2020} which allows user-processes
to make asynchronous I/O-related system calls.
Processes make system calls by adding an entry (SQE) to the \textit{submission
queue} (SQ), and receive the system call's return by polling the
\textit{completion queue} (CQ) for an entry (CQE).
A key option of \texttt{io\_uring} is a kernel SQ-polling thread which checks
for incoming SQEs from the process and sends the SQE work requests to a pool of
worker kernel threads---such a polling thread allows for a design where
system calls can be made without any traps in or out of the kernel.

\subsection{Making System Calls from an Enclave}
\begin{figure}[t]
	\centering
	\includegraphics[width=0.85\columnwidth]{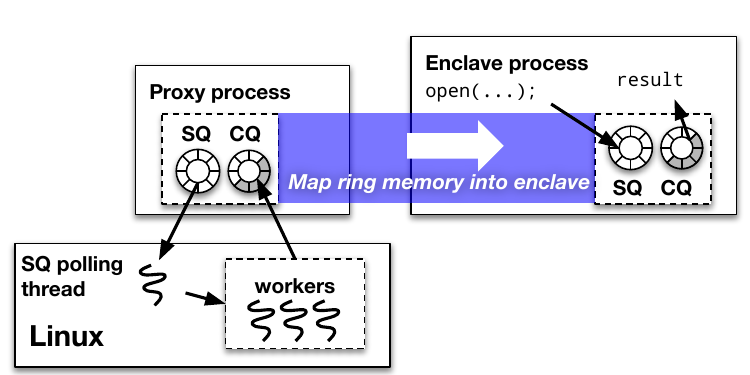}
	\caption{Illustration of how \texttt{io\_uring} SQ and CQ memory could be
	mapped into an enclave, giving it access to I/O}
	\label{fig:ring-mapping}
\end{figure}

For \ringleader, \texttt{io\_uring} provides an exciting opportunity to change
the hierarchical relationship between process and Linux kernel.
If a process is enclaved in the TrustZone, in a VM, or on a co-processor, and if
it could get access to SQ and CQ ring memory and shared I/O memory, then it
could request I/O ``system calls'' from Linux without needing to be directly
managed by Linux.
Continuing with the TrustZone example, if the SQ and CQ are mapped into a
TrustZone process, as shown in Fig. \ref{fig:ring-mapping}, then it could
perform arbitrary I/O system calls \textit{as if Linux was a remote file system
and I/O service}.
This is the basis of \ringleader's design.

For \ringleader, enclaves are processes running in the ARM TrustZone Secure
world or in an ARM Confidential Compute Architecture (CCA) Realm, \etc{}; they
are managed by a trusted OS, which handles scheduling, virtual memory, and other
essential OS tasks.
OP-TEE is popular open-source option, but we base our implementation and design
off of PARTEE \cite{PARTEE2025}, a real-time trusted OS.
We will call this updated TEE OS ``\ringleader OS'' in this paper.
As we will discuss further in \S \ref{section:enclave-availability}, this
is how enclaves can have liveness and non-starvation despite an untrusted
Normal world Linux OS.
\ringleader gives enclaves access to I/O system calls by mapping Normal-world
\texttt{io\_uring} SQ and CQ memory into an enclave; thus, enclave operations
on the queues are visible by the SQ-polling kernel thread.
All operations on the queue are non-blocking and lock-free, which lets enclave
execution continue regardless of the state of the queues.
While Linux could refuse to service system calls, it cannot block the execution
of a well-designed enclave.

\begin{figure}[t]
	\centering
	\includegraphics[width=0.80\columnwidth]{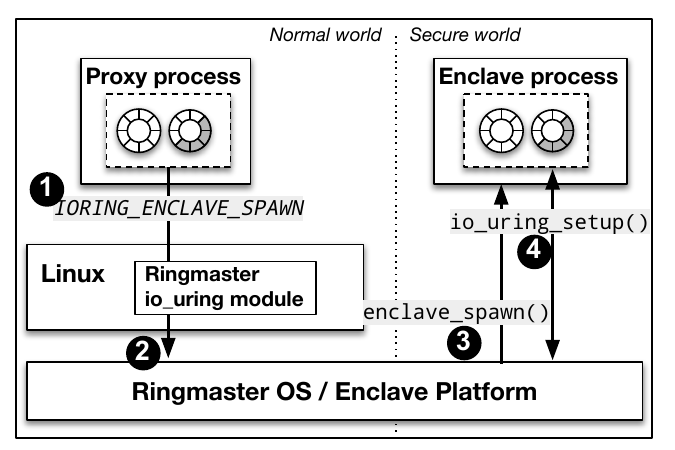}
	\caption{Diagram of \ringleader's process for registering and mapping
	\texttt{io\_uring} memory for an enclave}
	\label{fig:ring-registration}
\end{figure}

The safe mapping of \texttt{io\_uring}'s queues occurs through the following
steps, shown in Fig. \ref{fig:ring-registration}:
\begin{compactenum}

	\item Each enclave has a proxy process running as a Linux process; the
		proxy creates \texttt{io\_uring} queues, and then adds a
		\texttt{IORING\_OP\_ENCLAVE\_SPAWN} entry to the SQ.

	\item The \ringleader Linux kernel module implements this new SQE type.
        Once the SQE is received, Linux registers the queues' physical memory
        address with \ringleader OS.
		Additionally, Linux optionally notifies \ringleader OS to spawn a new
		enclave at this time, if it hasn't been started already.

	\item If the enclave isn't running, \ringleader OS spawns the enclave to be
		associated with the proxy by actually loading the process ELF binary
		into memory along with the arguments and environment variables.
        At this point \ringleader should optionally authenticate and attest the
        binary.

	\item The enclave checks for any registered \texttt{io\_uring} memory (via
		a synchronous system call serviced by \ringleader OS), and then
		\ringleader OS maps the SQ and CQ rings into the enclave's address
		space.
\end{compactenum}

\subsection{Shared Memory for Arguments}
\label{section:shared-memory}
Most system calls do not take simple value arguments, \eg{} in C/C++
\texttt{open()} requires a pointer to a file path string.
However, because \ringleader manages an enclave's address space, not Linux, then
a pointer in an SQE to the enclave's memory will not translate when read by
Linux.
Thus, \ringleader includes a two-part mechanism for establishing translatable
regions of shared memory between proxy and enclave.

\begin{figure}[t]
	\centering
	\includegraphics[width=0.90\columnwidth]{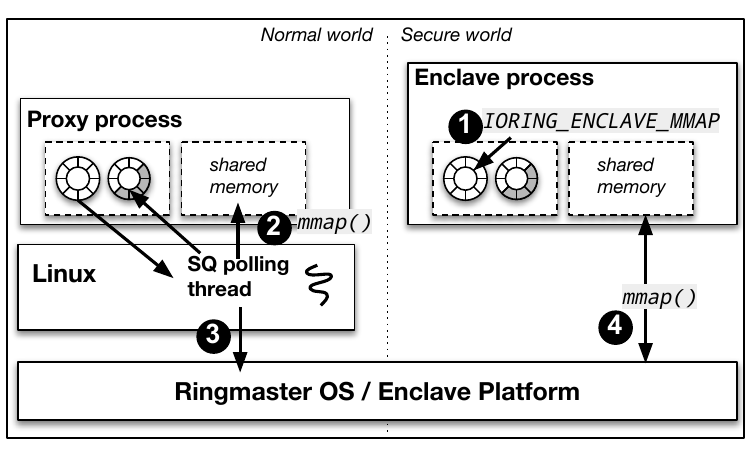}
	\caption{Diagram of \ringleader's process for registering and mapping
	generic shared memory for an enclave}
	\label{fig:shmem-registration}
\end{figure}

The first part of the mechanism is establishing shared memory between enclave
and proxy via the following registration process, shown in Fig.
\ref{fig:shmem-registration}:
\begin{compactenum}
	\item The enclave first adds a new \texttt{IORING\_OP\_ENCLAVE\_MMAP} entry
	to the SQ, specifying the request size with a identifier.

	\item The \ringleader module in Linux handles this new operation, which first
    	allocates a set of pages and then performs an \texttt{mmap} of the
    	requested size in the proxy.

	\item Linux then registers with \ringleader OS with the set of pages, and the
        identifier.
        At this point, it is finished, and it puts a return in the CQ; notably,
        the return value is the virtual address of the memory mapped into the
        \textit{proxy's address space}.

	\item Once the CQE is received, the enclave performs a \texttt{mmap} system
	call to \ringleader OS which then maps the registered shared memory into the
	enclave.
\end{compactenum}

\begin{figure}[t]
	\centering
	\includegraphics[width=1.0\columnwidth]{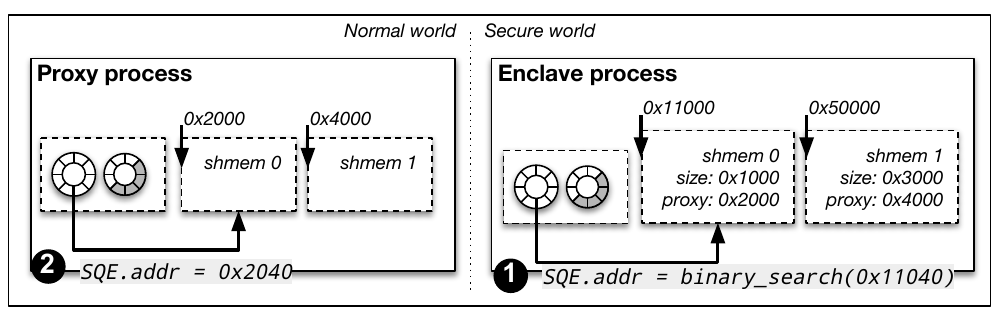}
	\caption{Diagram of \ringleader's process for translating pointers into shared
	memory where address \texttt{0x11040} in the enclave is translated to
	\texttt{0x2040} in the proxy's address space}
	\label{fig:shmem-ptr-translate}
\end{figure}

\noindent The second mechanism for passing pointers in SQEs is a transparent pointer
translation process.
\ringleader provides a set of user-space library functions to initialize each
SQE type with the correct system call arguments.
When an SQE is initialized, or ``prepped,'' with the arguments, \ringleader will
silently translate any pointer arguments that lie within the set of previously
mapped shared memory blocks.
To do this, \ringleader's user-space library maintains an ordered list of
shared memory blocks, with the enclave's address of the memory, the proxy's
address, and the block size---as shown in Fig. \ref{fig:shmem-ptr-translate}.
A binary search with the pointer argument on the list allows the pointer to be
translated from enclave to proxy.
For more complex data structures, \eg{} \texttt{writev} passes a pointer to an
array of pointers, a deep translation can be done of each pointer in the list.

With these mechanisms, an enclave already has everything it needs to perform
\texttt{io\_uring} operations through \ringleader.
To give the example of an \texttt{open} system call:
\begin{compactenum}

	\item First, acquire shared memory via an \texttt{IORING\_OP\_ENCLAVE\_MMAP}
	operation (or reuse old shared memory).

	\item Then, copy the file path into the shared memory.

	\item Next, enqueue an SQE with the path argument pointing to the shared memory
	   (internally \ringleader translates the pointer), and submit the SQE.

	\item The Linux SQ-polling thread will see the SQE in the
		proxy's queue, and will service the \texttt{open} system call; once it is
		complete the file descriptor will be created in the proxy process, and the
		descriptor number will be put in the CQ as the return value.

	\item The enclave now has opened the file and can use the descriptor number for
	further operations.
\end{compactenum}

\subsection{Power Management \& Wake-Up Signals}
\label{section:power}
Devices which use battery power or have limited thermal management typically
need to allow cores to sleep periodically, \eg{} using ARM's wait-for-interrupt
(\texttt{WFI}) instruction; however, this would not be possible if \ringleader
entirely relied on polling the \texttt{io\_uring} queues.
Normal systems manage SQ-polling thread's energy consumption impact by
setting a timeout, after which the thread will sleep.
Once this happens the program needs to use an \texttt{io\_uring\_enter()} system
call to wake up the thread again.
For \ringleader, once the thread sleeps the enclave would lose the ability to
make further \texttt{io\_uring} operations.
Furthermore, we cannot directly use Overshadow-style \cite{Overshadow2008,
TrustShadow2017} forwarding because we do not want to block the enclave waiting
for Linux.
Our solution is twofold: (1) Linux SQ-polling thread wake-up notifications, and
(2) leveraging a real-time scheduler.

We first designed a synchronous \texttt{io\_uring\_enter()} \ringleader OS
system call.
This system call makes \ringleader OS create a software-generated interrupt
(with some information about the calling enclave written into a shared memory
queue).
The interrupt will remain masked, however, so control will not immediately
transfer to Linux; instead, it will return back to the enclave until it yields
or its budget is exhausted.
Once Linux receives the interrupt, the \ringleader kernel module's handler will
now forward the system call on behalf of the enclave (using techniques described
by TrustShadow \cite{TrustShadow2017})---waking up the SQ-polling thread.

On the enclave side, we do not implement wake-up notifications; however, by
design, periodic tasks in a real-time scheduler go hand-in-hand with
polling-style I/O.
As we describe later in \S \ref{section:scheduling}, \ringleader uses a
real-time scheduler and enclaves can be configured with periods and budgets.
Thus, the pattern for proper power management would be, once the enclave wakes
each period: poll while there are any \texttt{io\_uring} completions, perform
work, create submissions, and then yield the remaining budget (\eg{} see Fig.
\ref{fig:event-loop-code}).
This common pattern real-time will allow unused budget to be used to either
sleep or perhaps run Linux.

\subsection{Handling Malicious Rich I/O}
\label{section:malicious-io}
As demonstrated by SCONE \cite{SCONE2016}, BlackBox \cite{BlackBox2022},
TrustShadow \cite{TrustShadow2017}, Occlum \cite{Occlum2020}, Graphene-SGX
\cite{Graphene-SGX2017}, and others, encryption and authentication of data
passed and returned by a system call can mitigate many Iago attacks
\cite{IagoAttacks2013} relating to the file system and network.

The remaining source of malicious system call data is from host-owned devices or
host user process---these should not be blindly trusted for safety-critical
operations.
We expect that \ringleader enclaves will always need to perform some safety
checks on such system calls, but this is already becoming common practice to
avoid spoofing attacks and robustness issues \cite{GhostInTheAir2012,
GPSSpoofing2014, IllusionAndDazzle2017, FoolingTheEyes2022, LongTail2019}.
Mitigation depends on the context; however, a benefit of \ringleader is that
enclaves are always given the opportunity to vet incoming data.
For LiDAR sensors, it may be possible to detect tampering with a watermarking
scheme \cite{LidarIntegrityVerification2019, SeeingIsNotAlwayBelieving2021,
LiDARWatermarking2024}.
The Simplex approach has also been explored in detail, to validate untrusted ML
outputs \cite{Simplex2001, UAVSimplex2016, SecureCore2013, VirtualDrone2017,
PROTC2017, Contego-TEE2019, RealTimeReachability2022, DynamicSimplex2023}.

\section{\ringleader Enclave API}
The \texttt{io\_uring} user-space library, \texttt{liburing} \cite{liburing},
implicitly trusts the OS in certain way, so we redesigned a new library which
considers an adversarial OS and potential programmer security pitfalls.
Practically, this means \ringleader must handle adversarial corruption of queues
while maintaining deterministic runtime, \ie{} we assume the contents of shared
memory are untrusted and can change at any point.
For example, \ringleader does not read the ring sizes from shared memory, except
to confirm that they are the expected size.
Additionally, the programmer must ideally have limited access to volatile shared
memory, and when they do, it should be clear.
In future work, static analysis could ensure at compile time that shared memory
operations are memory safe and will not lead to infinite loops.
Finally, we do not use any unbounded loops or locks in functions that involve
interfacing with ring memory, and we always bound the head and tail indices
stored in shared memory.
The next sections describe in more detail our designs.

\subsection{Dynamic Shared Memory Arenas}
\label{section:arena}
Because most \texttt{io\_uring} operations require pointers, and thus for a
\ringleader enclave require shared memory, the reuse and management of shared
memory is important for security, performance, and ease of programming.
While a generic memory management software pattern (\eg{} \texttt{malloc} /
\texttt{free}) would work to help allocate and manage shared memory, we observe
that shared memory allocation and freeing will likely follow the pattern of:
allocate memory for one or more objects, make one or more \texttt{io\_uring}
system calls, then free the memory.
Additionally many system calls have defined sequences, \eg{} \texttt{open},
\texttt{read}, \texttt{close}, or \texttt{socket}, \texttt{bind},
\texttt{listen}, \texttt{accept}; in these cases shared memory likely will not be
randomly interleaved, so a heavy-weight \texttt{malloc} for shared memory is
not always needed.

\begin{figure}[t]
	\centering
	\includegraphics[width=0.95\columnwidth]{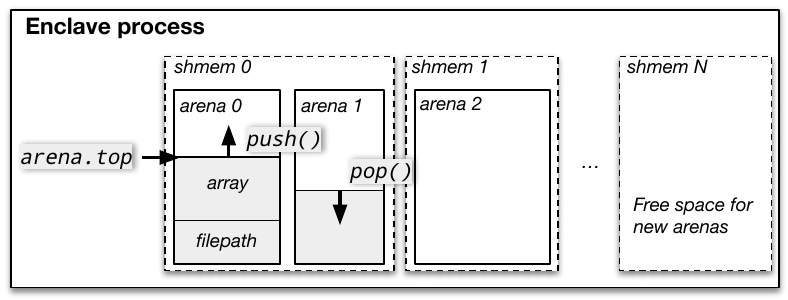}
	\caption{Diagram illustrating how \ringleader shared memory arenas contain
	objects with similar life cycles}
	\label{fig:ringleader-shmem-arena}
\end{figure}

With these patterns in mind, we chose an arena-based design
\cite{ArenaAllocation1990} for managing shared memory in a performant way.
\ringleader arenas are blocks of shared memory which can be used to quickly
allocate and free multiple objects that have a similar life cycle, throughout
the process of one or more related I/O operations.
We illustrate in Fig. \ref{fig:ringleader-shmem-arena} how multiple arenas can
be allocated from shared memory, and used to store objects associated with
\texttt{io\_uring} operations.

In \ringleader's arena-based design, an enclave requests a shared-memory arena
of a certain size.
\ringleader's user-space library immediately fulfills the request if shared
memory is available.
If no memory block is large enough, the library silently fills its pool of
shared memory with a \texttt{IORING\_OP\_ENCLAVE\_MMAP} request.
When the completion arrives from Linux, it can fulfill the arena request
asynchronously.

The power of an arena is through fast allocation and freeing of objects
within the arena's memory.
The constraint is that objects must be freed from an arena in
\textit{first-in-last-out}, stack order.
A push is a constant-time operation that allocates memory by incrementing the
\textit{top} of the arena's stack by the number of bytes in the allocation.
A pop decrements the top of the stack by $N$ bytes.
In this way, small objects, \eg{} file paths, socket address structures,
\etc{}, can be quickly allocated and freed with an associated arena once it has
been set up.
Further, objects do not even need to be popped in many cases, as they will all
be freed when the arena is freed.

\paragraph{Early request optimization:}
Given the potential overhead of the additional
\texttt{IORING\_OP\_ENCLAVE\_MMAP} calls, we want to minimize the number of
requests made, \ringleader has a few techniques for doing this (however, the
possibilities for optimizations are quite large here).
First, in many cases, we can request an approximation of the needed arena sizes
for an enclave even before the enclave's \texttt{main()} begins execution.
At enclave launch, \ringleader's user-space library checks for an environment
variable containing an initial request size for shared memory.
By making a large request immediately, the enclave can overlap the request time
with other initialization tasks.
Once the initial large block arrives, it can be divided into various
useful arena sizes, based on heuristics of common sizes.
By doing this we avoid potential overhead of the binary search used to translate
pointers.

\paragraph{Managing free arenas}
When it comes to managing freed arenas, we can borrow many techniques from
traditional memory allocators; however, the key difference is that we cannot
trust the contents of the shared memory blocks.
Traditional boundary-tag allocators typically have a block header which is
stored adjacent to the allocated or freed block.
For \ringleader, we want to avoid storing block metadata in shared memory
because an adversarial Linux could overwrite the contents of the header to
create a memory corruption exploit.
This means that block metadata must be stored in private enclave memory; and we
cannot use constant-time pointer arithmetic in the arena \texttt{free} function
identify the block's metadata.
The shared memory block metadata problem further reinforces our choice of an
arena-based system, where arena metadata is stored in a \texttt{struct} which
is owned by the programmer.
Once an arena is freed, the metadata struct can go into a free-list or binning
data structure, as used in dlmalloc \cite{dlmalloc1996}.

\subsection{Protections via Abstraction}

\paragraph{No direct access to the SQ and CQ}
We avoid giving the programmer direct access to SQ and CQ ring memory.
For example,
the \texttt{ringmaster\_try\_get\_sqe()} function, which is similar to
\texttt{io\_uring\_get\_sqe()}, does not return a pointer to the reserved SQE,
rather an index (of type \texttt{sqe\_id\_t} to the slot in the SQ.
We avoid giving the programmer direct access to CQEs in shared ring
memory; however, since CQEs are small, we opt to copy the data out of shared
memory into a private \texttt{struct io\_uring\_cqe}.
Thus the programmer gets access to a CQE pointer similar to \texttt{liburing}.

\paragraph{Protected \texttt{user\_data}}
In an \texttt{io\_uring} SQE, the \texttt{user\_data} field allows the
programmer to associate an identifier or data with an operation.
It is a simple 64-bit field passed into the SQE and returned in the CQE.
We observed that this field could be the source of many exploits, if the OS
manipulates it.
So \ringleader abstracts the true \texttt{user\_data} field internally and
provides a safe user-data field.
We store the programmer's \texttt{user\_data} in a private table, and use a
value internal to the \ringleader library which is bounds checked for safety.
Additionally, this allows \ringleader to mitigate double-completion attacks by
forcing a monotonically increasing internal field value.

\paragraph{Avoiding Infinite Loops}
Additionally, for this reason, we do not use any unbounded loops or locks in
functions that involve interfacing with ring memory, and we always bound the
head and tail indices stored in shared memory.
The process of prepping and submitting an SQE involves only straight-line code
with no loops or recursion; the key step is an
atomic increment of the tail value and an atomic read of the head value.
If there is no room in the SQ, the function returns an error code.
Likewise, getting a CQE also involves no loops or recursion, so run times are
theoretically deterministic outside of microarchitectural side-effects or page
faults.

\subsection{Promises Framework}
\label{section:promises}

After using \ringleader extensively to build different applications, it became
clear that it is very difficult for programmers to manage potentially long
chains of asynchronous events.
Further, C lacks language features that make asynchronous programming
easier and safer.
Thus, we needed to provide some solution to help programmers manage
asynchronous events easily and safely.

Thus we turned to a well-known programming abstraction (often used in JavaScript
for managing RESTful APIs), called \textit{promises}.
At a high level, the programmer can simply call \texttt{ringmaster\_read} or
\texttt{ringmaster\_write} which return promise objects.
The promise objects simplify a chain of asynchronous events and callback functor
objects (functions with arguments).
Inside \texttt{ringmaster\_write}, it requests a free shared memory, resulting
in another promise.
It then calls \texttt{ringmaster\_sqe}, which returns a promise for an SQE; this
could happen if the submission queue ring is full because Linux hasn't processed
the SQEs.
Then it chains onto that promise with one that waits for pending writes to be
completed, and so on.
At the end of the function, the \texttt{ringmaster\_write} function returns the
final promise in the chain.

Internally, \ringleader's promises store a function pointer, next pointer, and a
fixed size array of arguments.
Effectively, a promise call chain acts like an asynchronous call stack.
The only thing the programmer has to do is to poll for the promise to be
complete (or busy wait), thus vastly simplifying management of DoS attacks and
availability.%
\footnote{In the future, we plan to implement \ringleader as a user-space Rust
package, since Rust handles asynchronous events at the programming language
level this is a strong choice for new development.}
Promises are allocated in chunks dynamically, this means the OS could delay
system call returns so that promises build up and consume more and more memory.
To prevent memory DoS, we bound the promise call stack with a static
configuration variable.
As we will see later, the ability to buffer data while waiting on a promise
opens up a number of system call optimization features (\S
\ref{section:async-optimizations}).

\section{\ringleader LibC and LibOS: Legacy Applications}
\label{section:ringleader-libc}

We designed a \ringleader LibC (based on Musl LibC) which can be compiled with
unmodified source code to run as an enclave.
We modify the implementation of LibC functions to perform ring operations via
\ringleader instead of making synchronous I/O system calls.
Internally, after enqueuing an SQE (for example, when \texttt{read()} is
called), the LibC waits for the completion of that system call.
For system calls implemented by \ringleader OS, we can make them synchronously
as normal.
Table \ref{tab:syscall-table} also shows how we divided up the calls.

Similarly, we plan for a future version of \ringleader to implement a
hardware-abstraction layer for Gramine \cite{Graphene-SGX2017}, a library OS.
The result would be an even greater possibility of applications and languages
that could run with TrustZone protection---essentially as portable TrustZone
containers.

\begin{table*}[t]
	\begin{center}
		\footnotesize
		\begin{tabular}{| p{2in} | l | p{3.25in} |}
			\hline
			\textbf{Service / System Call} & \textbf{OS} & \textbf{Rationale} \\
			\hline

			\hline
			page-table management, \texttt{mmap},
			\texttt{munmap}, \texttt{mprotect}, \texttt{mremap}, \texttt{sbrk},
			\texttt{brk}, \etc{} & \ringleader & Difficult to protect against Iago
			attacks \cite{IagoAttacks2013}, private data can be leaked through page
			access patterns \cite{PreventingPageFaults2016} \\

			\hline
			scheduling, \texttt{yield()}, \texttt{gettimeofday()} & \ringleader
			& Necessary for availability.  \\

			\hline
			randomness & \ringleader & Linux-controlled randomness
			would allow cryptographic replay attacks \\

			\hline
			signal handling, \texttt{sigalarm()}, \etc{} & \ringleader & Needed
			for availability, requires fine-grained access to enclave code \\

			\hline
			critical device access, I2C, UART, SPI, CAN drivers, \etc{} &
			\ringleader & Needed for safety- or timing-critical I/O availability
			\\

			\hline
			pthreads, semaphores, mutexes, \etc{} &
			\ringleader & Required for availability (future work for prototype
			implementation) \\

			\hline
			\texttt{spawn()} & \ringleader & Necessary for enclave
			process-level parallelism \\

			\hline
			\texttt{getpid()}, \texttt{getppid()} &
			Linux / \texttt{io\_uring} & Useful for getting proxy's process ID,
			\\

			\hline
			file system operations, \texttt{open()}, \texttt{read()}, \texttt{write()},
			\texttt{close()}, \texttt{unlink()}, \texttt{mkdir()}, \texttt{statx()},
			\etc{} & Linux / \texttt{io\_uring} & Complex file system drivers are already
			available \\

			\hline
			\texttt{socket()}, \texttt{bind()}, \texttt{connect()}, \texttt{accept()},
			\texttt{recv()}, \texttt{send()}, \etc{} & Linux / \texttt{io\_uring} &
			Complex network drivers and TCP/IP stack drivers already available in Linux;
			most systems assume unreliable networking \\

			\hline
			\texttt{posix\_spawn()} & Linux / \texttt{io\_uring} & Can be
			supported with the addition of proposed \texttt{IORING\_OP\_CLONE}
			and \texttt{IORING\_OP\_EXEC} \cite{IOUringSpawn2022} (Future work
			for current implementation) \\

			\hline
			file-backed \texttt{mmap()} & Unsupported / Future Work & Requires
			page-fault handling in Linux, which would break availability \\

			\hline
			\texttt{fork()} & Unsupported / Future work  & Requires forking
			both enclave process and proxy process state, will complicate
			\ringleader OS design \cite{AForkInTheRoad2019} and inflate TCB \\

			\hline
		\end{tabular}
	\end{center}
	\caption{Table showing how operating system services are divided for
	\ringleader enclaves, using a synchronous trap into \ringleader OS or
	asynchronous communication with Linux via \texttt{io\_uring}}
	\label{tab:syscall-table}
\end{table*}

\subsection{Time-Sensitivity with a Synchronous API}
\label{section:timeouts}
\ringleader allows a programmer to make some changes to legacy POSIX-style
applications to prevent blocking attacks from Linux.
Instead of waiting for the completion event, the programmer can define a
call-specific timeout for any system call.
After the timeout, \texttt{errno} will be set to \texttt{EAGAIN} or
\texttt{ETIMEOUT} to indicate the timeout failure.
Additionally, the programmer can register a \texttt{SIGALRM} signal, which will
interrupt the spin-loop waiting for the completion.

\subsection{Asynchronous Optimizations}
\label{section:async-optimizations}
Similar to previous work \cite{Occlum2021}, we optimize \ringleader LibC by
buffering in shared memory and making asynchronous writes and reads.
Implementing this optimization requires care not to violate POSIX standards or
application behavior expectations when working with regular files, pipes, FIFOs,
sockets, devices, etc (\ringleader disables optimizations after identifying the
the descriptor types for pseudo files like \texttt{/dev/*}, \eg{}).

We use our promise library's \texttt{ringmaster\_write}, to buffer all incoming
writes in shared memory.
From our experiments with regular files, it is safest to only have one ongoing
write system call at the same time, to avoid challenges with failures; however,
we find that we can typically optimally buffer up to the size of half of the
system's last-level cache.
Furthermore, we write and read in file-system optimal block sizes (determined
through \texttt{stat}).
We observe that buffering and promises overhead increases the latency of a
single write; however, if ``pipelining'' writes happens, the enclave actually
runs faster than baseline, see Fig. \ref{section:performance}.

In the future we plan to optimize other calls as well, when the user makes a
\texttt{listen()} call, we can automatically also enqueue a \textit{multi-shot}
accept \texttt{io\_uring} operation.
Internally we queue incoming accepts as they arrive.
The result is that we can have some parallelization and also reduce the number
of \texttt{accept} operations a program out need to send to the Linux kernel.

\section{Details on \ringleader OS \& Linux}
Here we explain the design details of \ringleader OS in order to clarify how
\ringleader meets its security claims.

\subsection{Secure System Calls \& Devices}
Similar to early enclave work on Proxos \cite{Proxos2006}, the system call
interface is split (some calls go asynchronously into Linux and some go normally
into \ringleader OS).
However, for \ringleader, the split division ensures not only confidentiality,
and integrity, but also availability.
The full table of how OS services are divided for \ringleader is given in Table
\ref{tab:syscall-table}.

\subsection{Background: Preemption and Scheduling of Linux}
\label{section:scheduling}
Past work on real-time enclaves and TrustZone-assisted hypervisors has
demonstrated how to preempt and schedule Linux on ARM architectures
\cite{SafeG2010, LTZVisor2017, VOSYSMonitor2017, RT-TEE2022, PARTEE2025}; we use
these core ideas for \ringleader, and will briefly explain how we applied these
designs.
Typically ARM cores come with a separate Secure world timer device, which can
only be configured from \texttt{S-EL1} (secure kernel mode) or \texttt{EL3}
(firmware or monitor mode).
We configure the timer to periodically interrupt and the control
registers to trap into \texttt{EL3} for secure timer interrupts.
From \texttt{EL3} mode, \ringleader OS can take over, saving and restoring
state, switching to enclaves, \etc{}.

\ringleader OS implements a real-time scheduler which can support
fixed-priority (FP), earliest-deadline-first (EDF), among other algorithms.
The scheduler is budget-enforcing: each enclave process has a time budget which
is replenished at the enclave's defined period; once the budget is exhausted
the enclave will not be scheduled until replenishment.
Each Linux core is represented as a thread in the \ringleader OS scheduler, the
main difference is instead of restoring only user-space registers, \ringleader
OS must also switch to \texttt{EL3} to swap \texttt{EL1} registers for this
type of thread.

\subsection{Starting Enclaves with Resource Donation}
An adversary may try to launch many enclaves in order to exhaust memory or time
resources---the two primary finite resources in our system design.
These resources need to be distributed securely when enclaves are allowed to be
dynamically started.
\ringleader OS leverage's PARTEE's budget-enforcing scheduler, which ensures
that CPU resources are bounded.
Furthermore, with a \texttt{ulimit}-style physical memory quota, \ringleader OS
can ensure that memory is also bounded per enclave process.
When spawning a new enclave, \ringleader OS implements a resource donation
system.
A parent thread must donate some of its time budget and memory quota to the
child process.

\subsection{Changes To Linux}

The set of default operations provided by \texttt{io\_uring} is incomplete for
many programs that only use \texttt{io\_uring} for I/O; for example, there are
no operations at this time for \texttt{bind()} and \texttt{listen()}.
For this reason, we implemented ten \texttt{io\_uring} operations so far in the
Linux kernel.
We found that converting a system call to an \texttt{io\_uring} operation can
be relatively easy (we added the \texttt{sync()} operation in less than 10
minutes of engineering time).%
In future work we plan to make the set of operations dynamically extensible via
kernel module; this way \ringleader is not tied to specific kernel versions.

\subsection{Thread- \& Process-Level Parallelism}

\subsubsection{\texttt{pthread}}
\ringleader can support multi-threaded enclaves as long as the \ringleader OS
supports thread-level parallelism.
For future work \ringleader OS can provide an interface for donating budget to a
new thread.
Additionally, \ringleader OS should provide mutexes to prevent infinite blocking.

\subsubsection{\texttt{fork()}}
It is likely possible to implement \texttt{fork()} for \ringleader, the design
would be to first invoke a \texttt{fork()} in the proxy process, for example
through the addition of a \texttt{IORING\_OP\_FORK} or through a enclave-proxy
IPC command pipe.
Then it would be needed to implement \texttt{fork()} as a \ringleader OS system
call.

Unfortunately, there are some complications with simply forking in this way.
First, the shared memory would need to be renegotiated somehow, as the regions
would be shared between child and parent.
Second, the forked proxy would need a way to connect with the new enclave,
\ie{} inform \ringleader OS about the proxy's new physical pages associated
with the \texttt{io\_uring} SQ and CQ rings; however, the proxy will need some
kind of handle for that new enclave process which may be difficult to identify.
Finally, \texttt{fork()} will add complexity to \ringleader OS, and as
discussed in recent research, \texttt{fork()} should likely be deprecated
because of the problems of state duplication in the kernel
\cite{AForkInTheRoad2019}.

\subsubsection{\texttt{posix\_spawn()} and \texttt{io\_uring\_spawn()}}

Josh Triplett recently presented proposed operations for \texttt{io\_uring}:
\texttt{IORING\_OP\_CLONE} and \texttt{IORING\_OP\_EXEC}
\cite{IOUringSpawn2022}.
The power of these two operations in \texttt{io\_uring} is that when linked
into chains of operations with the \texttt{IOSQE\_IO\_LINK} and
\texttt{IOSQE\_IO\_HARDLINK} flags, a chain of \textit{clone}, followed by some
linked \textit{open}, \textit{close}, and other file operations, followed by an
\textit{exec} operation allows an entirely programmable spawning interface to
happen with zero context switches.
If released publicly, these features would allow \ringleader to support
spawning new Linux processes---which could be \ringleader proxies that start
new enclaves.
Furthermore, this would also allow the implementation of LibC's
\texttt{posix\_spawn} using \texttt{io\_uring\_spawn} as the backend.

\section{Security Analysis}
In this section, we first perform an analysis of \ringleader based on Game 1 and
2 defined in \S \ref{section:adversary-model}.
Then, we evaluate a drone application, discussed in \S
\ref{section:drone-motivation}, with \ringleader security.

\subsection{Case study: A highly-secure drone}
\label{section:drone-implementation}

We implemented the drone shown in Fig.
\ref{fig:ringleader-drone-overview}.
We constructed a quad-copter platform with a CubePilot Orange autopilot device
and a Raspberry Pi4b companion computer running a Linux 6.5 Buildroot OS.%
\footnote{The Pi4b does not have bus security needed for the TrustZone (\eg{} no
TZASC), so nested page tables must be used for memory isolation.}
The companion receives high-level objectives from a ground-control station via a
radio link.
On the companion computer, a path is continually planned between the objectives
and sent to the autopilot via MAVLink over serial \cite{MAVLinkSurvey2019}.
Additionally, a LiDAR device is used by the companion computer to collect data
and avoid obstacles.
The path planner adjusts the plan based on these readings.
An improved design could also use a camera and computer vision stack on Linux to
identify additional high-level objectives and stream video, but this was not
necessary for our prototype.

\begin{figure}
    \centering
    \includegraphics[width=1.0\linewidth]{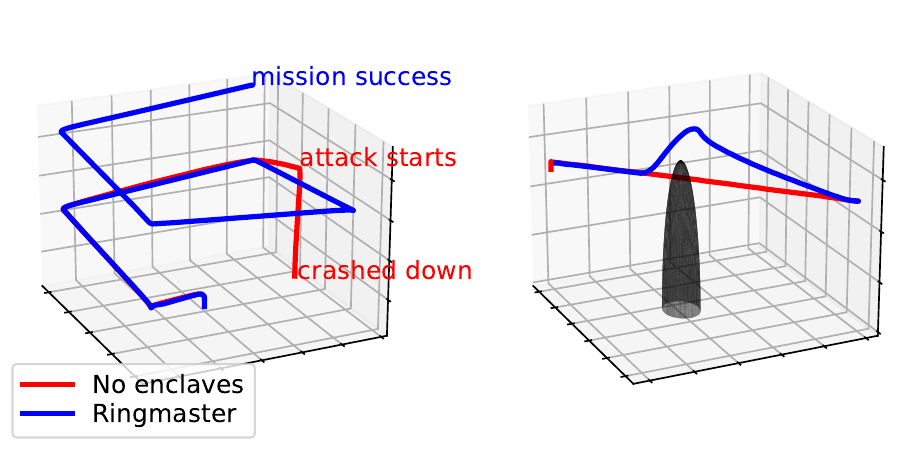}
    \caption{\textbf{Left:} Flight paths of same mission, showing how
    \ringleader protects the safety-critical MAVLink device from being attacked
    by the compromised host OS, prevent arbitrary crash; \textbf{Right:} Flight
    paths from the obstacle avoidance experiment, showing how an adversarial OS
    cannot affect the enclave's path-planner, so it avoids a near-miss
    scenario.}
    \label{fig:drone-trace}
\end{figure}

The MAVLink serial channel is safety critical, as the autopilot can be easily
manipulated.
We tested and confirmed that without any other measures, any companion computer
process can arbitrarily crash and control the drone by sending forced disarm or
manual control commands (see Fig. \ref{fig:drone-trace}).

While the autopilot handles real-time stabilization, the companion computer is
also highly time-sensitive because it must detect and avoid obstacles.
Thus, to validate the security of \ringleader we implement the path-planner
program so it can compile to run on \ringleader or as a regular Linux process.
The planner uses a TLS-encrypted connection with the ground station to ensure
that objectives are private and not corrupted by the OS.
Since the isolation of memory, device MMIO, and IRQ security are
well-understood hardware security features, we primarily focus our experimental
analysis on the exposed \ringleader interface.

\subsubsection{Obstacle-avoidance experiment:}
We performed an experiment where the path-planner enclave must read LiDAR data
to avoid an object collision (in this case just a ``near-miss''). The object is
not visible to the sensor at first, but once it appears in the flight path, the
planner should reroute; thus, the LiDAR sensor is safety-critical and will be
owned by \ringleader.
We tested two versions of this experiment: (1) no enclaves, \ie{} no
\ringleader protection, and (2) with \ringleader protections; results are shown
on the right in Fig. \ref{fig:drone-trace}.
For the former, we confirmed that if the adversary killed the planner process
after the mission started, the drone would fail the experiment.
For the latter, we were unable to make the drone ``hit'' the obstacle even with
killing the proxy, draining system resources, or crashing Linux entirely.

\subsection{Detailed Analysis}
\label{section:enclave-availability}

\begin{figure}[t]
	\begin{lstlisting}
while(1) {
  /* Handle all secure device events */
  while((res = chardev_read(serial_dev, buf, n)) > 0)
    handle_serial_data(res, buf, n);
  /* Handle up to max_sz io_uring events */
  for(int sz = 0; sz < max_sz; sz++) {
    struct io_uring_cqe * cqe = ringmaster_peek_cqe(rl);
    if(!cqe) break;
    res = ringmaster_cqe_get_result(cqe);
    switch(ringmaster_cqe_get_data64(cqe)) {
      case ARENA: handle_arena(cqe); break;
      case SOCKET: do_bind(res); break;
      case BIND: do_listen(); break;
      /* etc ... */
      default: handle_error(cqe); break;
    }
    ringmaster_consume_cqe(rl, cqe);
  }
  yield(); /* handled all events this period */
}\end{lstlisting}
	\caption{Example \ringleader enclave event loop}
	\label{fig:event-loop-code}
\end{figure}

In our adversary model's assumptions (\S\ref{section:adversary-model}, we
suggested that enclaves must be implemented to avoid potential timing
vulnerabilities.
However, this is not a major burden for enclave design; we illustrate this in
Fig. \ref{fig:event-loop-code} and discuss in \S \ref{section:promises}
This code is showing a very simple polling event loop, where an enclave is
handling incoming secure device and \texttt{io\_uring} events.
We can observe that by ``peeking'' on incoming completions, we can avoid hanging
waiting for the OS to respond.
With this in mind, we can analyze our \ringleader implementation against the two
games defined in our adversary mode.

\subsubsection{Game 1: Timeliness.}
Both the enclave and Linux will have a period and budget in \ringleader's
scheduler, so Linux will be preempted periodically after reaching its budget.
If it tries to power off the machine or change CPU frequency or voltage, the
firmware can be configured to deny access to these hardware features, as shown
in \cite{RT-TEE2022}. 
Multicore synchronization locks in \ringleader OS are MCS locks
\cite{MCSLock1991} with bounded wait times \cite{MCSLockLiveness2017}, and the
\ringleader OS kernel avoids recursive calls and uses only clearly bounded
loops.

All of the \ringleader API functions contain only bounded loops, so their
runtime has a theoretical upper bound.
If Linux refuses system call service, the enclave can still access the secure device.
If Linux corrupts the queue, the enclave will potentially receive junk values as
call returns---so authentication or bounds checking should be employed by the
program.
If Linux continuously fills the completion queue with returns, the \ringleader
library will internally ignore them as each completion must match a submission;
hence, the enclave cannot be delayed by a continual stream of completions.
The only exception is if multiple-completions are expected (\ie{} multi-shot
accept), in which case, the application needs to consider that peek may always
return non-null.
If Linux kills the proxy or even completely crashes, it will not affect the
execution of this event loop as the physical memory registered to the enclave
will remain mapped into its page tables.

Can Linux try to abuse the new \texttt{IORING\_OP\_ENCLAVE\_MMAP} or
\texttt{IORING\_OP\_ENCLAVE\_SPAWN} calls to DoS an enclave?
Since Linux can attempt SMC calls into \ringleader OS, it could try to
over-consume memory or \ringleader's data structures, or it could try to consume
CPU time by spawning many authentic enclaves repeatedly.
Since we based our \ringleader OS implementation on PARTEE \cite{PARTEE2025}, we
can ensure that Linux is bounded in its resource consumption.
This is because PARTEE partitions TEE OS data structures to prevent DoS attacks.
Additionally, once an enclave is started, PARTEE enforces its guaranteed CPU budget.

\subsubsection{Game 2: Confidentiality \& Integrity.}
To try to modify, or read, an enclave's private state (registers or data), Linux
would need to read or infer the non-shared memory pages belonging to the
enclave.
Linux cannot alter or read non-shared enclave physical memory directly since
\ringleader OS configures the memory bus to disallow reads or writes to the
physical address range from the Normal world.
Typically, this is a TrustZone Address Space Controller (TZASC) \cite{TZASC}
device (or the Granule Protection Table for ARMv9), or other SoC-specific
bus-security mechanism (nested paging must be used in the absence of bus
security).
This also prevents a DMA device from reading/writing enclave memory, as long as
the DMA device will be subject to the bus security or SMMU rules.
\ringleader OS manages enclave page faults, page tables, and memory allocation,
so traditional Iago attacks \cite{IagoAttacks2013} on \texttt{malloc} do not
apply.
The enclave's registers are stored in TrustZone-only memory on context switch.

In this game, another way a malicious OS can try to alter or read an
enclave's private state or communication is indirectly through manipulation of the
\texttt{io\_uring} interface.
We already discussed protections of the submission and completion queue shared
memory in Game 1, but an adversary could still corrupt system call return data.
As seen in Table \ref{tab:syscall-table}, the adversary has control over I/O
system calls, so \eg{}, it could manipulate a file read to send incorrectly
formatted data to confuse an application.
There are some solutions to these attacks, but, similar to past enclave system
call architectures \cite{Occlum2020, Graphene-SGX2017, TrustShadow2017}, we
assume application logic is implemented correctly and defensively, \eg{} error
conditions are always checked and I/O input is parsed securely.
The main defense against manipulation and snooping is encryption of files and
data, this can be done transparently if \ringleader is the backend of a Gramine
LibOS \cite{Graphene-SGX2017}.
For applications that access pseudo files (\eg{} \texttt{/dev/} and
\texttt{/proc/}), either it needs to be written defensively since the OS can
read arbitrary data into these buffers, or unmodified applications must use a
LibOS which emulates pseudo files.
One potential limitation is that the adversary could potentially infer private
state based on observing the patterns of system calls; this can be addressed
through related work via an ``oblivious'' file system \cite{Obliviate2018}.

The adversary can manipulate the two SMC calls that register \texttt{mmap}
shared memory or \texttt{io\_uring} params.
For example, it could try to allocate too little shared memory or shared memory
that overlaps with another enclave's private data.
This is prevented by design: because \ringleader OS confirms that there is no
overlapping in any regions of physical memory used for enclaves, the \ringleader
OS kernel, or device MMIO---this is done by maintaining a physical page
allocation table.
\ringleader enforces that the list of pages must all be unique and the total
size must match the equivalent to the size expected by the enclave (indicated by
the \texttt{mmap()} arguments).

\subsubsection{Reduced TCB.}
Currently for ARMv8, \ringleader OS is about 28k source lines of code (SLoC) in
C, and the user-space \ringleader library is only 1.2k SLoC.
Note that our OS also implements \texttt{EL3} functionality as well, so no extra
firmware is needed.
We provide an optional minimal LibC, which is 4.7k SLoC; the full \ringleader
LibC based on Musl (\S\ref{section:ringleader-libc}) is ~93k SLoC.
Linux is already nearly 40 million SLoC; thus, for systems with no other
protections, \ringleader already provides a massive reduction in TCB size.

\subsubsection{Limitations}
\ringleader is designed to provide availability and bounded execution for
time-sensitive enclaves while enabling access to rich OS services; however, it
does not address all security and systems challenges.
First, \ringleader does not address physical or microarchitectural side
channels.
We assume the adversary lacks physical access to the device, and that side
channels such as cache contention, memory bandwidth interference, or power
analysis are either out of scope or mitigated using complementary techniques.
Addressing such channels is orthogonal to \ringleader’s design.
Second, \ringleader does not guarantee availability of rich OS services. An
adversarial OS may delay, reorder, corrupt, or deny \texttt{io\_uring}-based
system calls entirely.
\ringleader guarantees that enclaves remain schedulable and responsive to secure
devices despite such behavior; correct handling of missing or malformed I/O data
remains application-specific.
Third, \ringleader assumes correctness of its TCB, including the \ringleader OS,
its real-time scheduler, and the hardware platform.
Vulnerabilities in these components are outside the scope of this work.
Finally, \ringleader currently requires kernel modifications to extend
\texttt{io\_uring} with enclave-specific operations.
While we found these changes to be modest, maintaining compatibility with
evolving Linux kernels may require engineering effort.
Future work could explore dynamically extensible or standardized interfaces to
reduce this burden.

\section{Performance Evaluation}
\label{section:performance}
\begin{table*}[t]
	\begin{center}
		\scriptsize
		
\begin{tabular}{|l | r r r | r r r || r | r | r | r | r | r |}
	\hline
	& \multicolumn{6}{c||}{Latency ($\mu$s)} & \multicolumn{6}{c|}{Overhead} \\
	\hline
	& \multicolumn{3}{c|}{Linux + GNU Libc} & \multicolumn{4}{c|}{\ringleader} & T.S.\cite{TrustShadow2017} & B.B.\cite{BlackBox2022} & V.G.\cite{VirtualGhost2014} & I.T.\cite{InkTag2013} & Proxos\cite{Proxos2006} \\
	\hline
	\textbf{Test} & min. & ave. & max. & min. & ave. & max. & ave. & rep. & rep. & rep. & rep. & rep. \\
	\hline 
	        null &       1.26 &       1.39 &      36.15 &       1.41 &       2.06 &      34.07 & \cellcolor[HTML]{ b8e593 }       1.48x & \cellcolor[HTML]{ dde593 }       2.01x & \cellcolor[HTML]{ e5db93 }       2.50x & \cellcolor[HTML]{ e5c093 }       3.90x & \cellcolor[HTML]{ e59793 }      55.80x & \cellcolor[HTML]{ e5a293 }      12.51x \\
	        open &       9.93 &      14.31 &     934.94 &      12.67 &      16.78 &     721.52 & \cellcolor[HTML]{ 94e593 }       1.17x & \cellcolor[HTML]{ afe593 }       1.40x & \cellcolor[HTML]{ bae593 }       1.50x & \cellcolor[HTML]{ e5b793 }       4.93x & \cellcolor[HTML]{ e5aa93 }       7.95x & \cellcolor[HTML]{ e59a93 }      32.61x \\
	        read &       1.48 &       1.63 &      84.02 &       2.37 &       3.02 &     585.04 & \cellcolor[HTML]{ d4e593 }       1.85x & - & \cellcolor[HTML]{ e1e593 }       2.10x & - & - & \cellcolor[HTML]{ e5a193 }      13.06x \\
	       write &       1.43 &       1.78 &      98.54 &       2.13 &       2.72 &     568.57 & \cellcolor[HTML]{ bde593 }       1.53x & - & \cellcolor[HTML]{ e1e593 }       2.10x & - & - & \cellcolor[HTML]{ e5a193 }      12.82x \\
	        stat &       4.19 &       4.74 &      44.02 &       6.63 &      12.93 &     586.37 & \cellcolor[HTML]{ e5d593 }       2.73x & - & \cellcolor[HTML]{ e5d193 }       2.90x & - & - & \cellcolor[HTML]{ e59e93 }      17.22x \\
	       fstat &       3.02 &       3.52 &      44.87 &       5.72 &      12.41 &     589.61 & \cellcolor[HTML]{ e5c593 }       3.53x & - & - & - & - & \cellcolor[HTML]{ e5a193 }      13.71x \\
	       mk 0k &      24.70 &      32.00 &     776.85 &      30.31 &      48.43 &     680.07 & \cellcolor[HTML]{ bbe593 }       1.51x & - & - & \cellcolor[HTML]{ e5b993 }       4.63x & - & - \\
	       rm 0k &      15.69 &      20.19 &     630.15 &      22.74 &      34.67 &     613.44 & \cellcolor[HTML]{ cbe593 }       1.72x & - & - & \cellcolor[HTML]{ e5b993 }       4.61x & - & - \\
	       mk 1k &      45.48 &      55.24 &     512.11 &      59.52 &      93.04 &     656.37 & \cellcolor[HTML]{ c9e593 }       1.68x & - & - & \cellcolor[HTML]{ e5b593 }       5.21x & - & - \\
	       rm 1k &      29.15 &      33.24 &     504.63 &      16.33 &      56.74 &    4115.24 & \cellcolor[HTML]{ cbe593 }       1.71x & - & - & \cellcolor[HTML]{ e5ba93 }       4.52x & - & - \\
	       mk 4k &      46.04 &      58.39 &     584.11 &      60.78 &      82.06 &     434.80 & \cellcolor[HTML]{ b0e593 }       1.41x & - & - & \cellcolor[HTML]{ e5b593 }       5.19x & - & - \\
	       rm 4k &      29.13 &      33.54 &     580.13 &      16.41 &      56.02 &     405.94 & \cellcolor[HTML]{ c8e593 }       1.67x & - & - & \cellcolor[HTML]{ e5ba93 }       4.52x & - & - \\
	      mk 10k &      58.57 &      74.37 &     658.63 &      74.02 &      97.67 &     700.02 & \cellcolor[HTML]{ a5e593 }       1.31x & - & - & \cellcolor[HTML]{ e5b993 }       4.71x & - & - \\
	      rm 10k &      36.09 &      43.44 &     646.15 &      26.89 &      66.81 &     403.96 & \cellcolor[HTML]{ bde593 }       1.54x & - & - & \cellcolor[HTML]{ e5b993 }       4.71x & - & - \\
	\hline
\end{tabular}

	\end{center}
	\caption{Table of LMBench Microbenchmarks, instrumented to run on \ringleader, overhead is compared with reported results from related enclave research}
	\label{tab:lat-table}
\end{table*}

To explore \ringleader's efficacy, we aim to answer the following
performance questions:
\textbf{Q1:} What is the I/O latency overhead of \ringleader?
\textbf{Q2:} What is the I/O throughput overhead of \ringleader?
\textbf{Q3:} What is non-I/O overhead of pure computation in an enclave?
\textbf{Q4:} What is the throughput of \ringleader when compared with an equivalent \texttt{io\_uring} application?
\textbf{Q5:} What is the overhead for unmodified programs using \ringleader's LibC?

A source of noise in performance experiments for \ringleader comes from the
choice of real-time scheduling parameters and how Linux chooses to schedule the
SQ-polling thread.
Thus, for these benchmarks we normalize execution time by dedicating one core
for enclave execution and giving the remaining cores 100\% utilization for
Linux; this way Linux can freely schedule its worker threads and SQ-polling
threads without interference noise in latency measurements.
Real-world system designers will have to tune the real-time parameters, core
affinities, \etc{} to their specific context.

\subsection{Latency Microbenchmarks}

We opted to test the latency of \ringleader by using a subset of the LMBench
tests \cite{LMBench1996} as this was a framework commonly used by other enclave
works.
We compiled LMBench sources with \ringleader LibC with minimal modifications so
that we can compare with existing works' reported overheads.
By doing this, our measurements also capture the overhead costs of (1)
\texttt{user\_data} protection, (2) arena allocation, (3) shared-memory address
translation, (4) copies into or out of shared memory.
We used as many tests as \ringleader supported, omitting any requiring
\texttt{fork()} and \texttt{exec()}.

Table \ref{tab:lat-table} shows the results of our experiments.
We observe comparable or better overhead for all benchmarks when compared the
reported results of with related enclave solutions: TrustShadow (T.S.)
\cite{TrustShadow2017}, BlackBox (B.B.) \cite{BlackBox2022}, Virtual Ghost
(V.G.) \cite{VirtualGhost2014}, and InkTag (I.T.) \cite{InkTag2013}.
The latency overheads of \ringleader are minimal due to the fact that we can
avoid the synchronous system call overhead and any extra work required in that
process by previous efforts (note that for BlackBox some of the overhead is due
to container protections as well).
However, the extra work required to do the four steps listed above, in addition
to multicore cache coherence, and possible overhead latency from
\texttt{io\_uring} does add at least a few hundred nanoseconds to each system
call.

\noindent\fbox{
    \parbox{\linewidth - 3\fboxsep}{%
		\textbf{Q1:} \ringleader has some latency overhead compared to
		synchronous Linux system calls, but it is comparable to, or better
		than, similar works.
    }%
}

\subsection{Throughput Comparison with \texttt{liburing}}
We designed benchmark applications which can be compiled into either a
\ringleader enclave, or a Linux process (using \texttt{liburing}).
These benchmarks measure throughput for parallelizable workloads and test how
the throughput changes when the application is ported to be an enclave.
We hypothesized that here \ringleader could achieve nearly zero overhead when
compared to a regular process because there are no significant differences in
the data paths, once enough shared memory has been established.
Table \ref{tab:fast-table} shows that in general we can receive large volumes of
data into a process very quickly---fully saturating the gigabit Ethernet port
of our Pi whether the code is running in an enclave or not.

\boxedtext{
	\textbf{Q2 \& Q4}: \ringleader has low observed overhead (0-3\%)
	compared to non-enclave \texttt{io\_uring} programs.
}

\begin{table}[t]
	\begin{center}
		\footnotesize
		
\begin{tabular}{|l | r | r | r |}
	\hline
	 Test & Linux/io\_uring & \ringleader & Overhead \\
	\hline
	read file &     946.79 MiB/s &     936.86 MiB/s &      1.011x \\
	write file &     826.41 MiB/s &     803.58 MiB/s &      1.028x \\
	TCP server &     111.37 MiB/s &     111.38 MiB/s &      1.000x \\
	\hline
\end{tabular}

	\end{center}
	\caption{Highly parallelized throughput tests for the network and
	file system, comparing a regular Linux progress to a \ringleader enclave}
	\label{tab:fast-table}
\end{table}

\subsection{Unmodified applications: GNU Coreutils, UnixBench}
\label{section:unmodified-performance}
\begin{figure}[t]
	\centering
	\includegraphics[width=0.95\columnwidth]{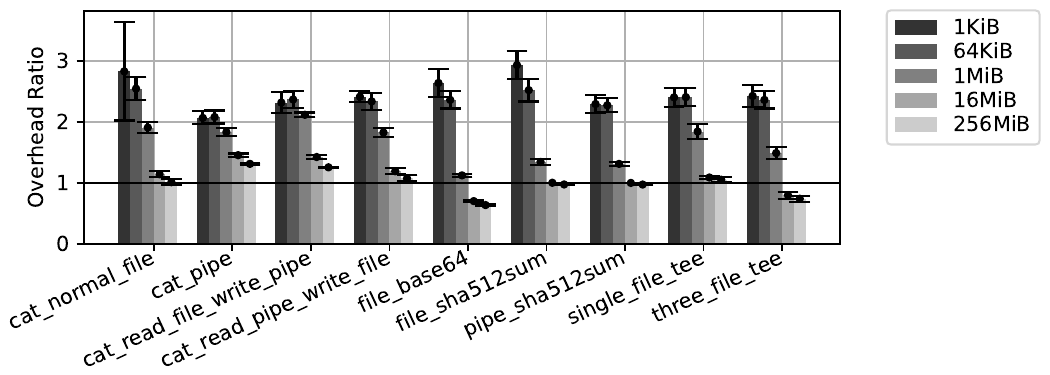}
	\caption{Benchmarks measuring overhead of unmodified GNU Coreutils programs
	linked against \ringleader's LibC.}
	\label{fig:Coreutils-performance}
\end{figure}

We tested \ringleader on complex real-world unmodified software by compiling
and linking a large portion (currently 22 programs) of the GNU Coreutils
programs with \ringleader LibC.
We chose a few programs which were amenable for more rigorous performance
analysis, \texttt{cat}, \texttt{tee}, \texttt{sha512sum}, and \texttt{base64},
and experimented with them.
We measured their overhead with three different file sizes, normalized against
the non-enclave version.

We observe the higher overheads for smaller files (shorter program runs).
This is due to extra startup costs associated with launching enclaves and
establishing shared memory (enclaves take about 12ms to start and compared to
about 5ms on average with our current approach).
For all tested programs programs we observe that we can get match or exceed
baseline performance with longer running operations.
In fact, the enclaved \texttt{base64}, \texttt{sha512}, and \texttt{tee}
applications had statistically significant performance improvements (nearly a
50\% increase for \texttt{base64}), due to the fact that they can utilize
parallelism from our buffered standard I/O implementation.

\boxedtext{
	\textbf{Q5}: \ringleader has some overhead costs for starting up, but
	approaches and can exceed baseline performance over time for the Coreutils
	applications we tested.
}

\begin{figure}[t]
	\centering
	\includegraphics[width=0.95\columnwidth]{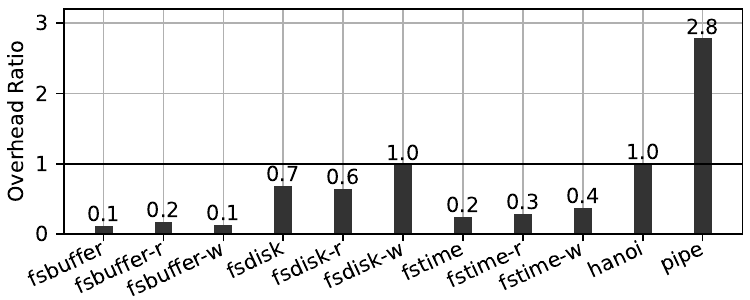}
	\caption{UnixBench Microbenchmarks, instrumented to run on \ringleader's LibC,
	compared against native Linux (lower is better).}
	\label{fig:unixbench-performance}
\end{figure}

We also measured throughput and raw computation overhead using UnixBench
compiled with \ringleader LibC.
We normalized \ringleader's scores against a standard execution of each test in
Fig. \ref{fig:unixbench-performance} (here lower numbers are better).
First, we can see that pure computation has no significant overhead with the
\texttt{hanoi} test.
The \texttt{fsbuffer} and \texttt{fstime} tests benefit significantly from
\ringleader's internal buffering, allow the system to batch multiple writes and
reads at the same time.
The \texttt{fsdisk} tests benefit less from buffering because the buffer usually
fills well before the disk I/O completes.
This is due to to the test's design, writing to a file that's larger than
Linux's internal file system buffers.
The \texttt{pipe} test, has the highest overhead we measured (2.8x).
This test is unable to benefit from any buffering optimizations, as each write
must fully complete before the next read.
Thus, the overhead is due to the extra shared memory copying, and the
\ringleader promises and \texttt{io\_uring} architectures.

\boxedtext{
	\textbf{Q2, Q3, \& Q5:} Throughput may be reduced for workloads which cannot
	benefit from I/O parallelism. We measured very low overhead for pure
	computation.
}

\section{Related Work}
\label{section:related-work}

\textbf{Asynchronous System Calls In SGX Enclaves:}
Solutions such as SCONE \cite{SCONE2016}, Eleos \cite{Eleos2017}, and Occlum
\cite{Occlum2020, Occlum2021} leverage asynchronous system calls and shared
memory to build SGX-based enclaves, but have three major challenges for
time-sensitive applications.

\textit{SGX shared memory is incompatible with other architectures:}
For SGX ``shared memory'' is any process data memory that is not enclave
memory, so allocating shared memory remains a simple \texttt{mmap()} or
\texttt{brk()} system call.
When in enclave mode, a program can still directly read and write
non-enclave memory, so these solutions were not faced with an additional
address-translation challenge (\S \ref{section:shared-memory}) that occur
outside of SGX-style enclaves.
Thus, \ringleader's innovation is a flexible and dynamic shared memory
design \textit{across address spaces and kernels}.

\textit{SGX is unable to provide availability:}
SGX is primarily for Intel (excepting HyperEnclave \cite{HyperEnclave2022})
which is less common for CPS, and we are not confident that SGX will ever be
able ensure real-time availability because it must rely on the untrusted OS
for scheduling and page-table management.

\textit{No asynchronous programming model or timeouts:}
Because these solutions focus on unmodified POSIX applications they do not
expose their asynchronous interface to the user.
Additionally, they do not require any innovations in how shared memory is
managed or exposed to users.
Because of the event-driven nature of many time-sensitive programs, we
anticipate that many \ringleader programs will benefit from its programming
model.
Even for unmodified modified programs, \ringleader LibC provides
timeouts and signals to ensure availability.

\textbf{Proxos:} Proxos \cite{Proxos2006} conceptually inspired \ringleader
by splitting the system-call interface between trusted and untrusted VMs.
However, Proxos lacks mechanisms to prevent infinite blocking on system calls
and did not support asynchronous system calls.
Additionally, its fixed-sized shared memory region and virtualization overheads
limit suitability for realistic workloads (see Table \ref{tab:lat-table}).

In theory, Proxos could be combined with FlexSC\cite{FlexSC2010}, SCONE
\cite{SCONE2016}, Eleos \cite{Eleos2017}, or Occlum \cite{Occlum2020,
Occlum2021}.
To the best of our knowledge, such a design has not been discussed or implemented
before, so \ringleader would be the first to explore this.
\ringleader also provides insights needed regarding dynamic shared memory,
availability via timeouts and signals, and power management.

\textbf{Unmodified Applications:}
When compared with other works supporting unmodified enclave designs
\cite{Overshadow2008, InkTag2013, Haven2014, Haven2015, SCONE2016,
Graphene-SGX2017, TrustShadow2017, Eleos2017, Panoply2017, Occlum2020,
Occlum2021, CHANCEL2021, BlackBox2022}, \ringleader's major unique contributions
are mechanisms to support availability (though with some modifications).
While next-gen Occlum \cite{Occlum2021} utilizes \texttt{io\_uring} to reduce
SGX switch overhead, it does not address real-time requirements or
inter-address-space memory sharing.
TrustShadow \cite{TrustShadow2017} forwards system calls from TrustZone to Linux
but cannot ensure availability due to its dependency on Overshadow-style page
table management \cite{Overshadow2008} even if it was extended with a timer
interrupt to wake tasks with timed-out system calls.

\textbf{Partitioning Hypervisors:}
Besides using a physically separate processor, partitioning hypervisors
\cite{StaticPartitioningHypervisors2023}, like Jailhouse \cite{Jailhouse2017},
PikeOS \cite{EvolutionOfPikeOS2007}, VxWorks \cite{VxWorks653}, Xen
\cite{Xen2003, RealTimeXen2014}, seL4 in SMACCM \cite{seL42009,
CanWePutTheSIntoIoT2022, SMACCM2017}, Bao \cite{Bao2020}, and the
TrustZone-assisted hypervisors \cite{SafeG2010, SASP2013, LTZVisor2017,
microRTZVisor2017, VOSYSMonitor2017} represent the dominant isolation technique
for CPS.
However, they have some challenging trade offs for real-world systems preventing
their widespread traction.
First, to communicate between the RTOS and general-purpose VM, they need a
VirtIO-like system \cite{TZ-VirtIO2018, RealTimeVirtio2021}.
This incurs significant engineering overhead to establish data flows, possibly
requiring the entire system hypervisor to be rebuilt for a single change to a
VM's task.
Secondly, virtualization simply for isolation may lead to excessive overheads
for some embedded systems \cite{EmbeddedHypervisorPerformance2016,
RoleOfVirtualization2008}.
Projects like Jailhouse trade the cost of VM exits for the usage of an entire
core, often 25-50\% of the processing power on SoCs.

\textbf{Secure I/O for Enclaves:}
LDR demonstrates how safely reuse Linux drivers in the TrustZone by dividing
their functionality between worlds \cite{LDR2024}; if paired with \ringleader,
this is a potential solution to the malicious I/O data problem.
Works like StrongBox \cite{StrongBox2022} and Graviton \cite{Graviton2018}
extend enclave isolation to GPUs, and also pair well with \ringleader.
StrongBox, for example, could securely receive encrypted GPU workloads via
\ringleader operations, enabling private AI/ML tasks.
MyTEE \cite{MyTEE2023} and RT-TEE \cite{RT-TEE2022} protect device access have
challenges scaling for broader applicability.

\section{Conclusion}
Modern safety-critical CPS have paradoxically conflicting needs: rich I/O to
meet consumer demands, and strong timing assurances for safety and reliability.
\ringleader strikes a delicate balance between these two by giving critical
software access to rich I/O while not forcing them to rely on it for safety
and security.
We presented new techniques which allowed protected enclaves to have access to
rich OS features while protecting it from many timing attacks.
Our results showed that an strong isolation plus an asynchronous approach to
rich I/O allowed an enclaves to have new timing guarantees which were previously
not possible---allowing for much greater security for robotics.

\bibliographystyle{acm}
\bibliography{ref}

\end{document}